\documentclass[pra,twocolumn, nofootinbib]{revtex4}

\usepackage{amssymb}
\usepackage{amsmath}
\usepackage{mathrsfs}
\usepackage{paralist}
\usepackage{graphicx}
\usepackage{xcolor}
\usepackage{footnote}

\newcommand{\ii}{{\rm i}}
\newcommand{\e}{{\rm e}}

\begin{document}
%
%
    \title{Zero reflection and transmission in graded index media}
%
%
    \author{C. G. King}
    \author{S. A. R. Horsley}
    \author{T. G. Philbin}
    \affiliation{Department of Physics and Astronomy, University of Exeter, Stocker Road, Exeter, EX4 4QL}
%
%
    \begin{abstract}
    Graded index media whose electric susceptibility satisfies the spatial Kramers-Kronig relations are known to be one-way reflectionless to electromagnetic radiation, for all angles of incidence. We demonstrate how a family of these media, in addition to being reflectionless, also have negligible transmission. To this end, we discuss how the transmission coefficient for the propagation of waves through a medium whose permittivity is built from poles in the complex position plane, with residues that sum to infinity, can be controlled by tuning the positions and residues of the poles. In particular, we have shown how to make the transmission  arbitrarily small, and hence maximise the absorption of the wave's energy. This behaviour is confirmed by numerical simulations.
    \end{abstract}
    \maketitle
%
%
    \section{Introduction}
When designing graded index media---where the electric permittivity varies continuously in space---it is useful to know which index profiles will completely absorb incident waves. Such materials (perfect absorbers (PAs)) totally suppress both reflection and transmission. Following the birth of transformation optics~\cite{Pendry2006,Leonhardt2006}, there has been a great deal of interest in finding materials which suppress the scattering of electromagnetic radiation without absorption, as these are an important ingredient in cloaking devices~\cite{Schurig2006}. Transformation optics can also be used to design PAs. For example, perfectly matched layers (PMLs)~\cite{Berenger1994} are commonly used in numerical simulations to absorb waves without reflection. However, they generally consist of an anisotropic permittivity and permeability, making them difficult to realise practically. In this work, we design non--reflecting, non-transmitting non-magnetic, isotropic graded index media, building on the recent findings on the reflectionlessness of spatial Kramers-Kronig media~\cite{Horsley2015,Horsley2016}. Absorbing graded-index media have been realised experimentally using metamaterials to create a Lorentzian shape permittivity profile (typical of a resonance in the frequency domain) whose permittivity satisfies the spatial Kramers-Kronig domain~\cite{Jiang2017}.
    \par
In order for electromagnetic radiation to be absorbed, the material must be lossy, meaning the permittivity is required to have a non-zero imaginary part. To design a PA, it may seem that one just needs to insert a slab of highly lossy material into the system. However, as the dissipation in the slab is increased, the reflection also increases until eventually nothing is absorbed.  Reflections at normal incidence can be reduced by impedance matching the slab with vacuum (equal relative permittivity and permeability, $\epsilon=\mu$) (see, for example~\cite{Stutzman2012}).  In contrast here we look to have low reflection for graded non-magnetic media for all angles of incidence where the dissipation is large and confined to a region of space on the order of the wavelength of incident radiation. Another means to achieve perfect absorption is by using coherent perfect absorbers (CPAs) or 'anti-lasers'~\cite{Chong2010,Wan2011}. Here waves are incident from opposite sides onto a lossy medium whose phase difference is such that the corresponding outgoing (or reflected) waves are eliminated through destructive interference, thus ensuring that all the waves energy ends up in the medium. Alternatively, thin layers of hyperbolic metamaterials have been shown to be perfectly absorbing to transverse magnetic (TM) polarised radiation in a relatively small frequency band~\cite{Nefedov2011}. Also metamaterials consisting of unit cells with two metallic resonators can be impedance matched to yield low reflection and high absorption in a small frequency band close to normal incidence~\cite{Landy2008}. Indeed metamaterials with gradually changing properties are one route to producing graded index media~\cite{Levy2004}. Such metamaterial perfect absorbers (MPAs) need not however be polarisation sensitive~\cite{Liu2016} and can additionally function for a wide range of frequencies and angles~\cite{Zhu2014}. A more comprehensive discussion of MPAs may be found in the review article~\cite{Ra'di2015}. In this work we explore a new way to determine distributions of dissipation in order to find media that do not reflect and totally absorb incident radiation.

    \section{Theory}
\begin{figure}[h!]
        \begin{center}
	\includegraphics[width=\linewidth]{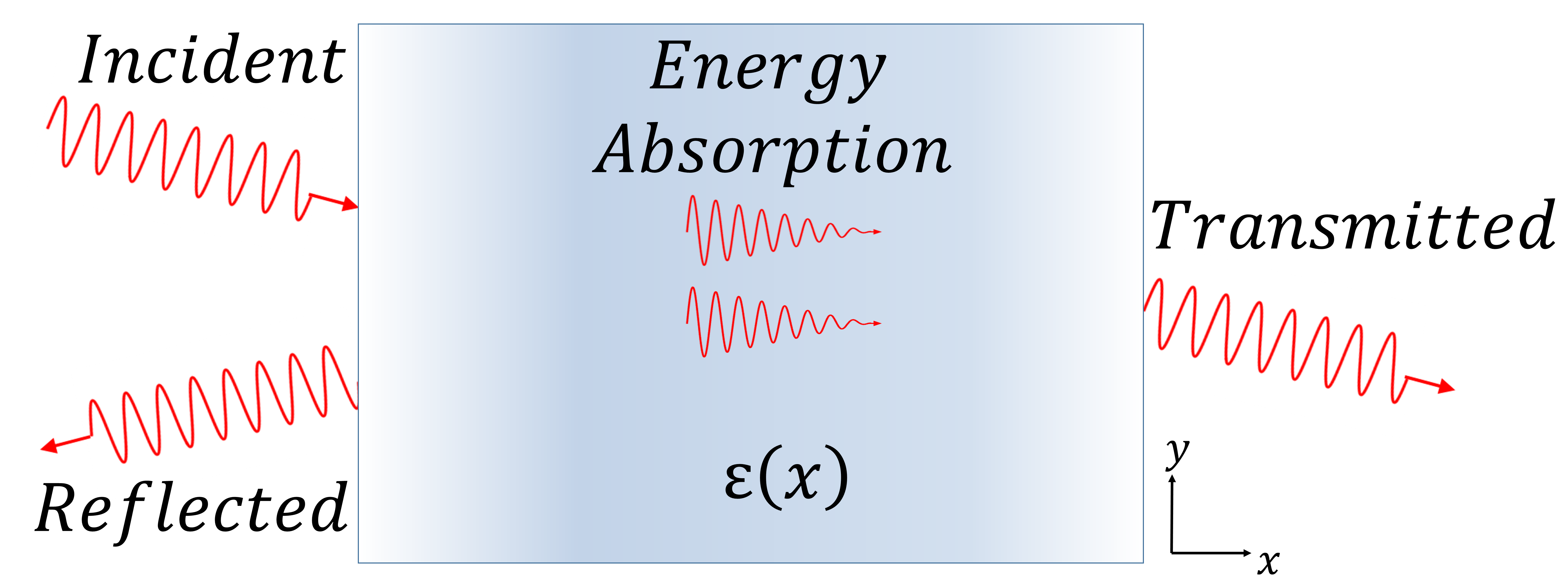}
        \caption{A wave of wavenumber \(k_{0}=\omega/c\) is incident from the negative \(x\) axis upon a material whose permittivity may be described by \(\epsilon(x)\) and is spatially homogeneous along the \(y\) direction. In general it is expected that the incident wave will split into three parts: a reflected wave, a transmitted wave and a part absorbed into the medium.\label{figure0}}
        \end{center}
    \end{figure}

    Consider the situation of a monochromatic wave of frequency \(\omega\) incident on a linear isotropic material whose permittivity varies only along one direction, as shown in figure~\ref{figure0}.
The electric field \(\varphi\) corresponding to a TE polarised plane wave propagating in the \((x,y)\)-plane through a material uniform in the \(y\) direction with permittivity \(\epsilon(x)\) and unit permeability satisfies the Helmholtz equation
    \begin{equation}
        \left[\frac{d^{2}}{dx^{2}}+k_{0}^{2}\epsilon(x)-k_{y}^{2}\right]\varphi(x)=0.\label{1}
    \end{equation}
Here \(k_{0}=\frac{\omega}{c}\) and \(k_{y}\) determines the angle of incidence of the wave. In what follows it is assumed without loss of generality that \(k_{y}=0\) \(^{1}\). \footnotetext[1]{Setting \(k_{y}=0\) corresponds to normal incidence. A non-zero \(k_{y}\) merely shifts the background value of the effective 1d permittivity.}Considering the situation of a medium sitting in free space, it is natural to split the permittivity up into its background value and the electric susceptibility \(\chi\) containing the inhomogeneous part \(\epsilon(x)=\epsilon_{\text{vac}}+\chi(x)\). Previous work has shown that specifying properties of the permittivity in the complex position plane \(z=x+\ii x'\) allows control of the reflection and transmission characteristics of a graded planar medium. Specifically graded index media whose susceptibility satisfies the spatial Kramers-Kronig relations
    \begin{equation}
    \begin{split}
        \text{Re}(\chi(x))&=\frac{1}{\pi}\mathbb{P}\int_{-\infty}^{\infty}\frac{\text{Im}(\chi(\hat{x}))}{\hat{x}-x}d\hat{x}\\
        \text{Im}(\chi(x))&=-\frac{1}{\pi}\mathbb{P}\int_{-\infty}^{\infty}\frac{\text{Re}(\chi(\hat{x}))}{\hat{x}-x}d\hat{x}\label{2}
    \end{split}
    \end{equation}
are reflectionless to waves incident from the left \(x=-\infty\)~\cite{Horsley2015,Horsley2016}. Such a condition corresponds to the permittivity being an analytic function of position \(z\) in the upper half complex position plane (\(x'>0\)), and decaying to \(\epsilon_{\text{vac}}\) as \(|z|\to\infty\) in this half plane.
    \par
The media we will be considering will decay as O\((1/x)\) or slower as \(x\to\pm\infty\) and the usual definition for the reflection and transmission coefficients for such extended profiles based on the incident, reflected and transmitted plane waves is not well defined due to the reliance on a reference point, as pointed out by Longhi~\cite{Longhi2015}, although reflectionlessness is still well-defined~\cite{Horsley2016(2)}. The moduli of the reflection and transmission coefficients can be defined unambiguously using the corresponding WKB waves~\cite{Heading2013}
    \begin{equation}
        \frac{1}{\epsilon(x)^{1/4}}\e^{\pm\ii k_{0}\int_{a}^{x}\sqrt{\epsilon(x')}dx'}.\label{2.1}
    \end{equation}
For spatial Kramers-Kronig media the transmission coefficient can be calculated as an integral in the upper half  complex position plane
    \begin{equation}
        |t|=\e^{k_{0}\text{Im}\left(\int_{C}\sqrt{\epsilon(z)}dz\right)},\label{3}
    \end{equation}
where the contour \(C\) goes from \(x=+\infty\) to \(x=-\infty\) along a path above any branch cuts of \(\sqrt{\epsilon}\)~\cite{Horsley2016} and this is valid for all angles and both left-to-right and right-to-left propagation (although there will be reflections from the right in general). Now consider a subset of the class of spatial Kramers-Kronig media: permittivity profiles containing a finite number of poles in the lower half position plane:
    \begin{equation}
        \epsilon(z)=1+\sum_{k=1}^{n}\frac{a_{k}}{z-z_{k}}+\sum_{k=1}^{m}\frac{b_{k}}{(z-p_{k})^{2}}+...\label{4}
    \end{equation}
Integrating around a large semi-circle \(C\) in the upper half plane contributes half a residue from each of the simple poles leading to a transmission coefficient~\cite{Horsley2016}
    \begin{equation}
        |t|=\e^{\frac{1}{2}\pi k_{0}\text{Re}\sum_{k=1}^{n}a_{k}}.\label{5}
    \end{equation}
It is therefore clear how to make graded index permittivity profiles having any desired transmission coefficient between 0 and 1 by tuning the residues of simple poles in \(\epsilon\). Also, by including only double or higher order poles, we can obtain profiles which give perfect transmission without reflection (somewhat surprisingly they can be used to design perfectly transmitting disordered media~\cite{King2017}). However, it is not clear how to make the transmission coefficient negligible without taking the limit of one of the simple pole residues going to \(-\infty\) (and hence the imaginary part of the permittivity to \(\infty\)). In the rest of this work, we explore how to make the transmission coefficient negligible whilst keeping the permittivity bounded.
    \par
High absorption has been achieved experimentally by a metamaterial having a profile of the form (\ref{4}) where the number of poles \(n\) is finite~\cite{Jiang2017}. However, for these spatial Kramers-Kronig media, to make the transmission coefficient completely vanish and have all of the incident wave absorbed, it follows from (\ref{5}) that we require a profile where the sum of the residues is infinite
    \begin{equation}
        \sum_{k}\text{Re}(a_{k})=-\infty.\label{6}
    \end{equation}
This requires an infinite number of simple poles in the lower half plane. For the series defining the permittivity (\ref{4}) to converge, and hence for \(\epsilon\) to be everywhere bounded, the sum \(\sum_{k=1}^{\infty}\frac{a_{k}}{z_{k}}\) must converge. In particular, there cannot be an infinite number of poles in a compact space of the complex plane whilst still satisfying condition (\ref{6}) and yielding a convergent permittivity (\ref{4}). Therefore, there must be poles extending infinitely far out in the lower half complex plane. In going from (\ref{3}) to (\ref{5}) for the transmission through a permittivity profile given by (\ref{4}), it was assumed that the number of poles, \(n\), is finite. However, it is not obvious that this holds in the limit \(n\to\infty\) i.e. when there are an infinite number of poles, because an interchange of limits is required: the limit as the radius of the contour \(C\) in (\ref{4}) tends to infinity must be swapped with the infinite summation(s) in (\ref{5}). Fortunately, with an application of Lebesgue's dominated convergence theorem~\cite{Bartle1995}, it can be shown that the transmission coefficient must vanish when the condition (\ref{6}) is satisfied.
    \par
The effect of having such an infinite sum of residues is to slow down the decay rate of the permittivity as \(x\to\pm\infty\). In particular, in the absence of any singularities in the permittivity \(\epsilon(x)\) along the real axis \(-\infty<x<\infty\), and any gain in the material (i.e. the permittivity has non-negative imaginary part), it is the speed of decay of the profile that determines whether the transmission coefficient is zero, unity or somewhere in between. This leads to the following classification:
    \begin{equation}
    \begin{split}
        |t|=1\qquad&\text{if }\chi(x)<\text{O}(1/x)\text{ as }x\to\infty\\
        0<|t|<1\qquad&\text{if }\chi(x)=\text{O}(1/x)\text{ as }x\to\infty\\
        |t|=0\qquad&\text{if }\chi(x)>\text{O}(1/x)\text{ as }x\to\infty.\label{7}
    \end{split}
    \end{equation}
Note that we are not saying anything about the real and imaginary parts of \(\chi\) separately. It is possible that they decay at different rates. In particular, it is possible for Im\(\chi\) to decay faster than \(\chi\), thus leading to an absorption higher than one would expect based on the decay of the lossy (imaginary) part of the permittivity.

    \section{Examples}
\begin{figure}[h!]
        \begin{center}
	\includegraphics[width=\linewidth]{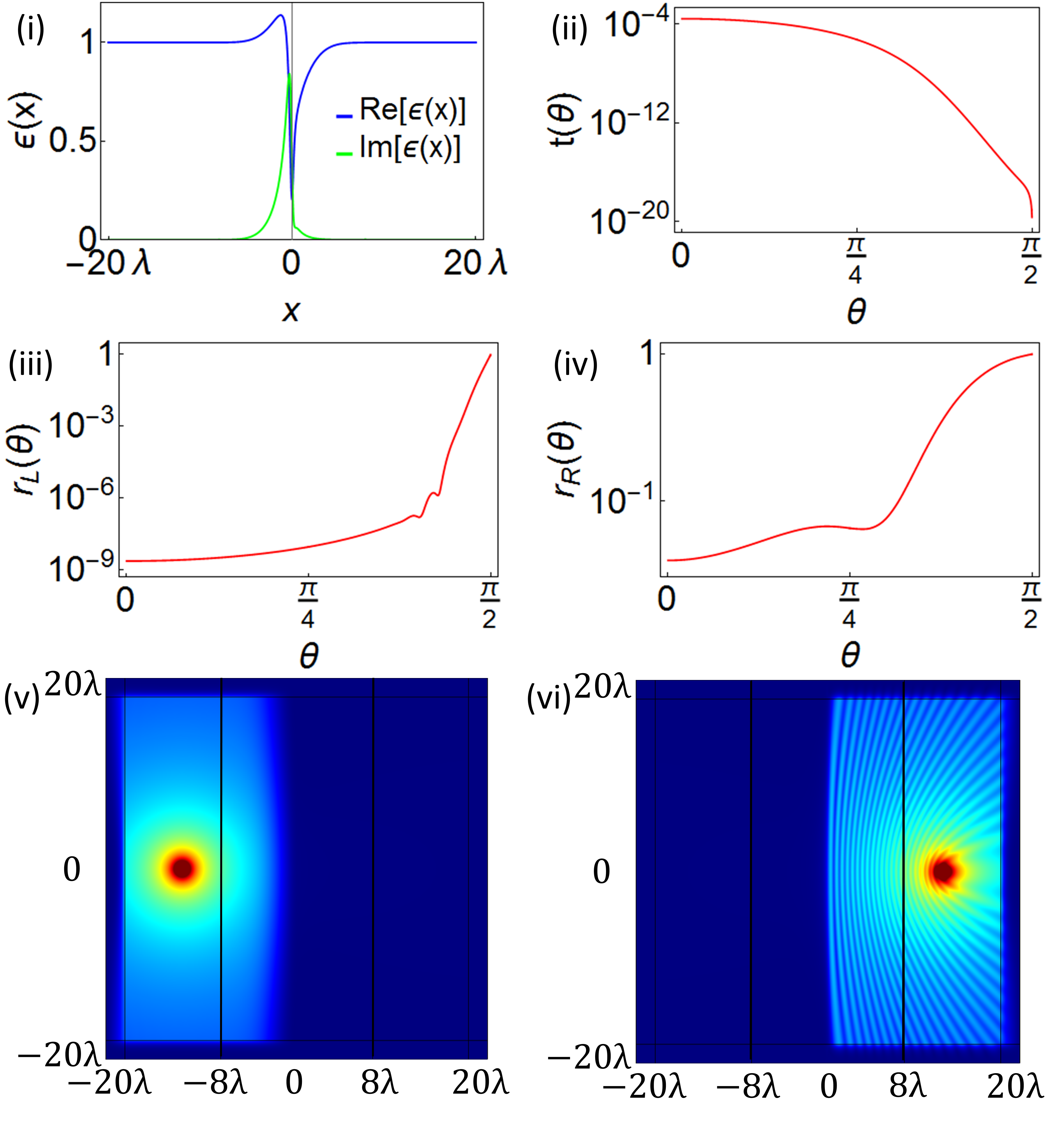}
        \caption{(i) The real and imaginary parts of the permittivity profile \(\epsilon(x)=1-\frac{\text{ln}\left(k_{0}x+2\ii\right)}{k_{0}x+2\ii}\), truncated at \(x=\pm50/k_{0}\). (ii) The transmission coefficient on a logarithmic scale of a plane wave propagating in either direction through the medium as a function of incidence angle. (iii) and (iv) The reflection coefficient for left and right incidence, respectively, as a function of incidence angle. (v) and (vi) Wave propagation simulated in Comsol Multiphysics~\cite{Comsol} by placing an out-of-plane line source to the (ii) left and (iii) right of the medium and the time-average of the absolute value of the electric field is plotted. Reflection and transmission is suppressed for a left incident wave whereas just the transmission is suppressed for a right incident wave.\label{figure1}}
        \end{center}
    \end{figure}

    As an example of this, consider taking \(a_{k}=-\frac{\alpha}{k}\) and \(z_{k}=-\beta k\ii\) where \(\alpha\) and \(\beta\) are positive constants. This corresponds to the profile
    \begin{equation}
        \epsilon(x)=1-\frac{\alpha}{x}\left(\gamma+\psi\left(1-\frac{\ii x}{\beta}\right)\right),\label{8}
    \end{equation}
where \(\psi(x)\) is the digamma function (the logarithmic derivative of the \(\Gamma\) function) and \(\gamma\) is Euler's constant. Having constructed this function from an infinite number of simple poles, note that asymptotically 
    \begin{equation}
        \epsilon(x)=1-\frac{\alpha\text{ln}x}{x}+\text{O}\left(\frac{1}{x}\right)\qquad\text{as }x\to\pm\infty,\label{9}
    \end{equation}
so here it is the weaker ln\((x)/x\) decay which leads to a profile with zero transmission. Since all terms beyond the second term do not affect the transmission (see (\ref{4}) and (\ref{5})), for simplicity consider the profile obtained by neglecting the O\((1/x)\) part of (\ref{9}) and displacing the remaining singularity to the lower half plane (which merely causes an O\((1/x)\) alteration in the profile):
    \begin{equation}
        \epsilon(x)=1-\frac{\alpha\text{ln}(\frac{x}{a}+\delta\ii)}{\frac{x}{a}+\delta\ii},\label{10}
    \end{equation}
where \(\delta\) and \(a\) are positive constants. This preserves the required property of being analytic in the upper half complex position plane, therefore satisfying the spatial Kramers-Kronig relations. Wave propagaion through such a material is simulated in figure~\ref{figure1}
where the material has been spatially truncated on either side at \(\pm8\lambda\). The region where there is significant dissipation is relatively small so we can consider these slabs as effectively thick coatings of non-absorbing material that supress the reflection from the thin region of large dissipation. This allows for spatial truncation of the profile without affecting the reflectionlessness of the material. Approaching grazing incidence the reflection becomes significant since imperfections in the truncation of the profile are magnified.
    \par
Note that, perhaps surprisingly, as \(x\to\infty\), the imaginary part decays as O\((1/x)\) in this example (as opposed to O\([\text{ln}x/x]\)), so we are in the situation described at the end of the theory section whereby the susceptibility's imaginary part decays faster than the full complex susceptibility. From the lossy part of the medium one might expect a transmission coefficient akin to that of a simple pole i.e. between 0 and 1. However, it is the interplay between the real and imaginary parts of the permittivity that not only combines to give zero reflection, but also the zero transmission property.
    \par
\begin{figure}[h!]
        \begin{center}
	\includegraphics[width=\linewidth]{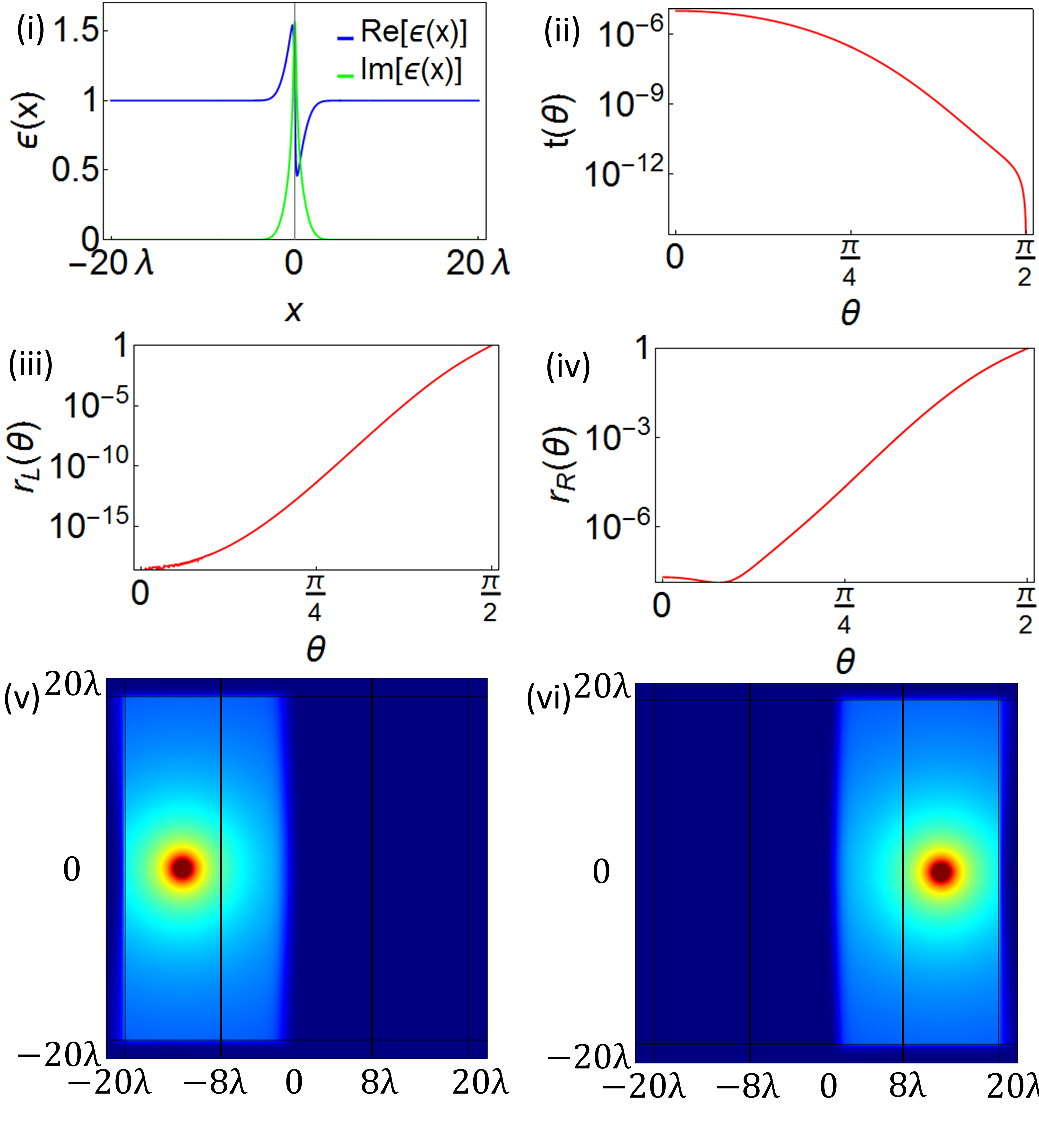}
        \caption{(i) The real and imaginary parts of the permittivity profile \(\epsilon(x)=1-\frac{\e^{-\pi\ii/4}\pi}{2\sqrt{k_{0}x+\ii}}\), truncated at \(x=\pm50/k_{0}\). (ii), (iii) and (iv) The transmission, left incidence reflection and right incidence reflection coefficient of a plane wave propagating as a function of incidence angle. (v) and (vi) Wave propagation simulated by placing an out-of-plane line source to the (ii) left and (iii) right of the medium and the time-average of the absolute value of the electric field is plotted. Reflection and transmission is suppressed for a left incident wave.\label{figure2}}
        \end{center}
    \end{figure}

As an example of a slightly different behaviour of decay for the permittivity which still gives zero reflection and transmission, consider taking \(a_{k}=-\alpha\) and \(z_{k}=-\beta k^{2}\ii\) where \(\alpha\) and \(\beta\) are positive. This corresponds to the permittivity
    \begin{equation}
    \begin{split}
        \epsilon(x)&=1-\frac{\alpha}{2x}\left(-1+\frac{\e^{\frac{\pi\ii}{4}}\pi\sqrt{x}}{\sqrt{\beta}}\text{cot}\left(\frac{\e^{\frac{\pi\ii}{4}}\pi\sqrt{x}}{\sqrt{\beta}}\right)\right)\\
        &=1-\frac{\e^{-\pi\ii/4}\alpha\pi}{2\sqrt{\beta x}}+\text{O}\left(\frac{1}{x}\right)\qquad\text{as }x\to\pm\infty.\label{11}
    \end{split}
    \end{equation}
Following the same process of neglecting the O\((1/x)\) part of the expansion leads to the simpler permittivity profile (again with the singularity displaced to the lower half plane to preserve analyticity in the upper half plane):
    \begin{equation}
        \epsilon(x)=1-\frac{\e^{-\pi\ii/4}\pi}{2\sqrt{\frac{x}{a}+\delta\ii}}\label{12}
    \end{equation}
where \(a\) and \(\delta\) are positive constants. In this case the decay is again slower than O\((1/x)\); in this case it is O\((1/\sqrt{x})\) for both the real and imaginary parts. This is plotted in figure~\ref{figure2}, (again truncated to \(16\lambda\)).
Much of the behaviour for this example is similar to the previous example. However, in this case, it is noticeable that the reflection is also low for a right incident wave. Reflections from general 'bumps' in graded index profiles are difficult to quantify a priori, though this example may open the possibility of suppressing transmission and reflection from both sides of the medium.
    \par
One of the advantages of these perfectly absorbing spatial Kramers-Kronig media are their capability to perform at all angles of incidence and at different frequencies. The profiles discussed in this paper have a thickness (the region in which Im(\(\epsilon\)) is noticably different from zero) on the order of a wavelength O\((\lambda)\). A thin (subwavelength) absorber can therefore be obtained by scaling the profiles appropriately:
    \begin{equation}
    \begin{split}
        \epsilon(x)&=1-\frac{\text{ln}\left(Ak_{0}x+2\ii\right)}{Ak_{0}x+2\ii}\\
        \epsilon(x)&=1-\frac{\e^{-\pi\ii/4}\pi}{2\sqrt{Ak_{0}x+\ii}},\label{13}
    \end{split}
    \end{equation}
where \(A>1\). Such a scaled-down profile may be better suited practically as an absorber for coating objects. However, one does have to bear in mind that, by imposing this scaling, the amount of absorption will then be greater restricted by the truncation, particularly at higher angles of incidence. Ultimately, the thickness required depends on the degree of imperfection one is happy with.
    \par
As discussed earlier, it is possible to create graded index media based on metamaterials~\cite{Levy2004}, where the material properties (i.e. the permittivity) can be tuned by gradually changing the individual resonant elements comprising the medium on a scale smaller than the wavelength. Indeed this has been used as a route to experimentally realising spatial Kramers-Kronig media~\cite{Jiang2017} and is thus probably the best known current route to obtaining a physical realisation of the profiles discussed in this work.

\section{Conclusion}
We have used the method of design developed in previous work~\cite{Horsley2015,Horsley2016} to examine permittivity profiles satisfying the spatial Kramers-Kronig relations where the transmission and reflection are zero. The earlier work has explained why reflection vanishes for Kramers-Kronig media but hasn't considered how the transmission can additionally be removed. The decay of the profiles can be classified into three types on the basis of the transmission properties, (faster than \(1/z\), full transmission; as \(1/z\), partial transmission; slower than \(1/z\), zero transmission). A typical graded index absorber has a permittivity with an imaginary part increasing slowly in space. We have designed perfectly absorbing permittivity profiles, localised to a central region of the profile, which decay to zero at infinity, but are constructed from sums of poles with infinite net residues. They have negligible transmission for all angles of incidence and negligible reflection away from grazing incidence. The findings were numerically verified for a finite slab of material using Comsol Multiphysics~\cite{Comsol}.

\subsection{Acknowledgements}
CGK acknowledges financial support from the EPSRC Centre for Doctoral Training in Electromagnetic Metamaterials EP/L015331/1. SARH acknowledges financial support from  EPSRC program grant EP/I034548/1, the Royal Society and TATA. TGP acknowledges financial support from EPSRC program grant EP/I034548/1.


\begin{thebibliography}{99}
        \bibitem{Pendry2006} J. B. Pendry, D. Schurig and D. R. Smith, \emph{Science} \textbf{312}, 5514, (2006).
        \bibitem{Leonhardt2006} U. Leonhardt, \emph{Science} \textbf{312}, 5781, (2006).
        \bibitem{Schurig2006} D. Schurig, J. J. Mock, B. J. Justice, S. A. Cummer, J. B. Pendry, A. F. Starr and D. R. Smith, \emph{Science} \textbf{314}, 5801, (2006).
        \bibitem{Berenger1994} J. Berenger, \emph{J. Comp. Phys.} \textbf{114}, 185, (1994).
        \bibitem{Horsley2015} S. A. R. Horsley, M. Artoni and G. C. La Rocca, \emph{Nature Phot.} \textbf{9} 436 (2015).
        \bibitem{Horsley2016} S. A. R. Horsley, C. G. King and T. G. Philbin, \emph{J. Opt.} \textbf{18} 4 (2016).
        \bibitem{Jiang2017} W. Jiang, Y. Ma, J. Yuan, Ge Yin, W. Wu and S. He, \emph{Laser and Phot. Rev.} \textbf{11} 1 (2017).
        \bibitem{Stutzman2012} W. L. Stutzman and G. A. Thiele., 2012 \emph{Antenna Theory and Design} (John Wiley and Sons).
        \bibitem{Chong2010} Y. Chong, Li Ge, H. Cao and A. D. Stone, \emph{Phys. Rev. Lett.} \textbf{105}, 053901, (2010).
        \bibitem{Wan2011} W. Wan, Y. Chong, Li Ge, H. Noh, A. D. Stone and H. Cao, \emph{Science} \textbf{331}, 889, (2011).
        \bibitem{Nefedov2011} I. S. Nefedov, C. A. Valagiannopoulos and L. A. Melnikov, \emph{J. Optics} \textbf{15}, 114003, (2011).
        \bibitem{Landy2008} N. I. Landy, S. Sajuyigbe, J. J. Mock, D. R. Smith and W. J. Padilla, \emph{Phys. Rev. Let.} \textbf{100}, 207402, (2008).
        \bibitem{Levy2004} U. Levy, M. Nezhad, H.-C. Kim, C.-H. Tsai, L. Pang and Y. Fainman., \emph{J. Optics} \textbf{22}, 4, (2005).
        \bibitem{Liu2016} X. Liu, Ke Bi, Bo Li, Q. Zhao and Ji Zhou, \emph{Optics Express} \textbf{24}, 18, (2016).
        \bibitem{Zhu2014} J. Zhu et. al., \emph{App. Phys. Lett.} \textbf{105}, 021102, (2014).
        \bibitem{Ra'di2015} Y. Ra'di, C. R. Simovski and S. A. Tretyakov, \emph{Phys. Rev. App.} \textbf{3}, 037001, (2015).
        \bibitem{Longhi2015} S. Longhi, \emph{Eur. Phys. Lett.} \textbf{112} 64001 (2015).
        \bibitem{Horsley2016(2)} S. A. R. Horsley, M. Artoni and G. C. La Rocca, \emph{Phys. Rev. A} \textbf{94} 063810 (2016).
        \bibitem{Heading2013} J. Heading, \emph{An Introduction to Phase-Integral Methods} Dover (2013).
        \bibitem{King2017} C. G. King, S. A. R. Horsley and T. G. Philbin, \emph{Phys. Rev. Lett.} \textbf{118} 163201 (2017).
        \bibitem{Bartle1995} R. G. Bartle, \emph{The elements of integration and Lebesgue measure} Wiley (1995).
        \bibitem{Comsol} COMSOL Multiphysics® v. 5.1. www.comsol.com. COMSOL AB, Stockholm, Sweden.
    \end{thebibliography}
\end{document}